\documentclass[useAMS]{mn2e}
\usepackage{graphicx}
\usepackage{epsfig}
\usepackage{amssymb}
\usepackage{threeparttable}
\usepackage{url}

\def\ltsima{$\; \buildrel < \over \sim \;$}
\def\lsim{\lower.5ex\hbox{\ltsima}}
\def\gtsima{$\; \buildrel > \over \sim \;$}
\def\gsim{\lower.5ex\hbox{\gtsima}}

\newcommand{\be}{\begin{equation}}
\newcommand{\en}{\end{equation}}
\newcommand{\ergs}{\rm \ erg \; s^{-1}}

\def\msole {~M_{\odot}}

\def\deg {$^\circ$}
\def\nh{\hbox{$N_{\rm H}$}}

\begin{document}
\title[]{A year in the life of the low-mass X--ray transient Aql X-1}

\author[Coti Zelati, Campana, D'Avanzo \& Melandri]{F. Coti Zelati$^{1,2}$\thanks{E-mail: francesco.cotizelati@brera.inaf.it}, 
S. Campana$^{1}$, P. D'Avanzo$^{1}$, A. Melandri$^{1}$ \\
$^1$ INAF -- Osservatorio astronomico di Brera, Via Bianchi 46, I--23807 Merate (LC), Italy\\
$^2$ Universit\`a di Milano-Bicocca, Dipartimento di Fisica G. Occhialini, Piazza della Scienza 3, I--20126 Milano, Italy
}

\maketitle

\begin{abstract}
The {\it Swift} satellite monitored the quiescence of the low-mass
X--ray binary transient Aql X-1 on a weekly basis during the March-November
2012 interval. A total of 42 observations were carried out in the soft
X--ray (0.3--10 keV) band with the X--ray Telescope on board {\it Swift}. 
We investigated the X--ray variability properties of Aql X-1 during 
quiescence by tracking luminosity variations and characterising them with a 
detailed spectral analysis. The source is highly variable in this phase and two
bright flares were detected, with peak luminosities of $\sim4\times10^{34}\ergs$ 
(0.3--10 keV). Quiescent X--ray spectra require both a soft thermal component 
below $\sim$2 keV and a hard component (a power law tail) above $\sim$2 keV. 
Changes in the power law normalisation alone can account for the overall observed 
variability. Therefore, based on our data set, the quiescent X--ray emission of
Aql X-1 is consistent with the cooling of the neutron star core and with 
mechanisms involving the accretion of matter onto the neutron star surface or magnetosphere.  
\end{abstract}


\section{Introduction}
Low-mass X--ray binaries (LMXBs) are among the brightest sources of the X--ray 
sky. They are binary systems composed by a compact object -- a neutron star (NS)
or a black hole (BH) -- and a late-type companion star with mass $\lsim1\msole$. 
The mechanism responsible for the X--ray emission in LMXBs is the accretion of 
matter onto the compact object, which occurs through Roche lobe overflow: the
companion star, filling its own Roche lobe, transfers gas to the Roche lobe of 
the compact object, forming an accretion disc (Frank, King \& Raine 2002).

Some X--ray binaries are persistent, with a constant luminosity of 
$10^{36}$--$10^{38}\ergs$ in the X--ray band. Others are instead transient
(low-mass X--ray transients, LMXTs), namely, they alternate periods of low 
luminosity (quiescence), to periods of high luminosity (outbursts). The 
former lasting years and with X--ray luminosities of order $10^{31}$--$10^{33}\ergs$, 
the latter lasting weeks or months and with peak luminosities comparable to 
the ones of the persistent sources. The currently accepted model to explain 
the transient behaviour is the thermal-viscous disc instability model 
(Lasota 2001), according to which the disc oscillates between a cold 
neutral state (i.e. quiescence) and a hot ionised state (i.e. outburst). 
Matter builds up in the disc during quiescence and is then transferred to the
compact object during outburst. 

Observationally, quiescent X--ray spectra of LMXTs hosting a NS (NS-LMXTs)
usually consist of a soft thermal component below $\sim$2 keV and a hard 
component above $\sim$2 keV (Campana et al. 1998; Rutledge et al. 1999; 
Campana 2004). The soft component is well described by pure hydrogen atmosphere 
models (e.g. Zavlin, Pavlov \& Shibanov 1996; Heinke et al. 2006) and it is usually 
interpreted as thermal emission from the NS surface in the cooling phase 
following an outburst. According to the deep crustal heating model (Brown, 
Bildsten \& Rutledge 1998; Colpi et al. 2001), matter accreted in outburst 
compresses the innermost regions of the crust, thus triggering nuclear reactions
and heating the core. The heat released is then radiated thermally in the 
subsequent quiescent phase. Emission from crustal cooling should decrease 
monotonically with time on time-scales of years, while emission due to core 
cooling is expected to be much more stable in time. Moreover, if residual 
accretion onto the NS is occuring in quiescence, a thermal like X--ray 
spectrum should be produced, due to the heating of the surface (Zampieri et al. 1995). 

The hard component is well represented by a power law tail with a photon index 
$\Gamma\sim1-2$, but its physical nature remains still debated. Among the 
suggested explanations, accretion onto the magnetospheric boundary seems to 
be an intriguing possibility. Although the physics of the disc-magnetosphere 
interaction is not well understood at low accretion rates (i.e. in the 
propeller regime), it is generally accepted that the disc is truncated near 
the compact object, where the magnetic field is more intense. The standard 
paradigm for the propeller regime (Illarionov \& Sunyaev 1975) predicts that
accreting matter might be pushed outward by the centrigugal drag exerted by 
the rotating magnetosphere. However, recent studies demonstrate that if the
magnetospheric radius is not far away from the corotation radius, the 
disc-magnetosphere interaction will not supply enough energy to expel matter 
from the disc and matter will instead accrete onto the magnetosphere (``dead
disc solution''; D'Angelo \& Spruit 2010, 2012). Accumulation of matter at the 
magnetospheric boundary can give rise to an inward pressure that might push 
the magnetospheric radius inside the corotation radius, thus allowing matter 
to bypass the centrifugal barrier and accrete onto the neutron star surface. 
Furthermore, the interaction between
the disc and the magnetic field in the propeller regime can lead to the 
formation of a hot, comptonising corona in the inner disc, which results in a
spectrum with a power law shape. 

For lower accretion rates, the NS can activate as a radio pulsar and the
interaction between matter flowing from the companion and the relativistic 
wind from the pulsar can cause a shock emission. A power law like spectrum 
extending from the optical to the X--ray band is predicted (Tavani \& Arons 
1997). Recently, the system PSR J1824$-$2452I in the globular cluster M28
was found to swing between being a radio millisecond pulsar and an accreting  
X--ray pulsar and viceversa, suggesting that these phases can alternate over 
a time-scale of a few weeks (Papitto et al. 2013).

It is not well known if, and how, the soft and hard spectral components are 
related. In the last 15 years, variability in the quiescent emission of
several NS-LMXTs has been observed on various time-scales (from hundreds of 
seconds up to years) and different physical mechanisms have been proposed in 
order to describe this variability (e.g. Campana et al. 1997, 2004; 
Rutledge et al. 2002; Campana \& Stella 2003; Cackett et al. 2005, 2011; 
Fridriksson et al. 2011; Degenaar \& Wijnands 2012; Wijnands \& Degenaar 2013; 
Bernardini et al. 2013).

The {\it Swift} satellite (Gehrels et al. 2004) monitored the quiescence of the 
NS-LMXT Aql X-1 during the March--November 2012 interval. A total of 42 observations
were carried out both in the soft X--ray (0.3--10 keV) and in the UV/optical bands 
with the X--ray Telescope (XRT) and the Ultraviolet and Optical Telescope (UVOT) on 
board {\it Swift}, respectively. In this paper we investigate the quiescent properties
of Aql X-1 with the aim to physically constrain the powering mechanism(s) at work 
during this phase. In Section 2 we summarise the main characteristics of Aql X-1.
In Section 3 we describe the {\it Swift} observations and the data reduction. In 
Section 4 we focus on the data analysis, discussing our results in Section 5.

\section{The source: Aql X-1}
With more than 40 outbursts observed in the X--ray and/or optical bands since its 
discovery in 1965 (Friedman, Byrom \& Chubb 1967), Aql X-1 is the most prolific 
X--ray transient known to date (20 outbursts were detected in the 1996--2011 epoch; 
Campana, Coti Zelati \& D'Avanzo 2013). Observations of type I X--ray bursts lead to
identify the compact object of this system with a neutron star (Koyama et al. 1981). 
The optical counterpart is known to be of late-K spectral type, with a quiescent magnitude
of $21.6\pm0.1$ mag in the $V$ band (Chevalier et al. 1999). An interloper 
located only $0.48''$ east of the true counterpart heavily complicates the studies 
in the quiescent state (Chevalier et al. 1999; Hynes \& Robinson 2012). The distance 
and the orbital period are estimated to be 5 kpc (Jonker \& Nelemans 2004) and $18.97$ 
h (Welsh, Robinson \& Young 2000), respectively. Casella et al. (2008) reported the 
discovery of a coherent millisecond X-ray pulsation in this source, lasting $\gsim150$ s 
and with a frequency of $550.27$ Hz. This value is very close to the maximum reported 
frequency from type I burst oscillations observed during an outburst in February-March 
1997 (Zhang et al. 1998) and corresponds to a spin period of $1.82$ ms. 

Aql X-1 has been monitored several times in its quiescent state. Rutledge et al. 
(2002) analysed three {\it Chandra} observations of this source relative to the
quiescence following the November 2000 outburst. They discovered X--ray variability, 
which they interpreted in terms of variations of the neutron star effective 
temperature, and found short--term ($<10^4$ s) variability in the 0.5--10 keV range 
during the last observation. The authors suggested that a variable mass accretion 
rate in the quiescent state is responsible for the observed variability. 
A different interpretation of the same data has been given by Campana \& Stella 
(2003): differences between spectra can be accounted for by correlated variations 
of the column density (which they interpreted as changes in the amount of mass 
outflowing from the companion) and the power law component, while the thermal component
remains constant. This variability can be explained in terms of shock emission that 
arises due to the interaction between the radio pulsar relativistic wind and matter 
flowing from the companion. Cackett et al. (2011) reported 14 archival observations of 
Aql X-1 (11 with {\it Chandra}, 3 with {\it XMM-Newton}), covering a time interval of 
$\sim$2 yr (November 2000--October 2002) and concerning quiescent epochs following 
different outbursts. They found clear variability between the observations. In particular,
a significant flare was observed in June 2002, with an increase in the flux by a factor 
of $\sim$5. The authors could not conclude whether the thermal component and/or the power 
law is responsible for the variability, but suggested that accretion at low rates during 
quiescence may be a likely explanation.

\section{Observations and data reduction}
{\it Swift} observed Aql X-1 in quiescence for 42 times in 2012 between March 15 and November 9, 
more or less once per week. A summary of the XRT observations is given in Table 1. 

\begin{table*}
\caption{Quiescent observations of Aql X-1 with {\it Swift}/XRT.}
\begin{center}
\begin{tabular}{cccccc}
\hline
Obs. ID     	& Date (2012)	& Exp. time	& Net count rate		& Number of counts	& Background percentage	\\
		& mm/dd 	& (ks)		& (10$^{-2}$ c s$^{-1}$)	&			& (\%)			\\
\hline
00031766040	& 03/15		& 4.8		& $6.2 \pm 0.4$			& 239			& 1.7\\
00031766041	& 03/21		& 5.1		& $4.6 \pm 0.3$			& 199			& 2.2\\
00031766042	& 03/27		& 3.5		& $10.6 \pm 0.6$		& 322			& 1.3\\
00031766043	& 04/02		& 3.9		& $7.4 \pm 0.5$			& 238			& 1.2\\
00031766044	& 04/08		& 5.2		& $5.1 \pm 0.4$			& 213			& 1.4\\
00031766045	& 04/15		& 5.1		& $3.2 \pm 0.3$			& 122			& 1.2\\
00031766046	& 04/20		& 5.1		& $2.3 \pm 0.2$			& 88			& 1.5\\
00031766047	& 04/26		& 4.7		& $3.3 \pm 0.3$			& 120			& 1.0\\
00031766048	& 05/02		& 5.2		& $2.5 \pm 0.3$			& 98			& 1.7\\
00031766049	& 05/08		& 3.4		& $2.8 \pm 0.3$			& 73			& 1.1\\
00031766050	& 05/14		& 3.4		& $2.9 \pm 0.3$			& 73			& 1.2\\
00031766051	& 05/20		& 4.9		& $3.4 \pm 0.3$			& 117			& 1.1\\
00031766052	& 05/26		& 3.0		& $1.5 \pm 0.3$			& 34			& 2.3\\
00031766053	& 05/28		& 1.6		& $2.2 \pm 0.4$			& 29			& 1.4\\
00031766054	& 06/01		& 4.7		& $2.5 \pm 0.3$			& 91			& 1.5\\
00031766055	& 06/08		& 4.8		& $1.8 \pm 0.2$			& 63			& 2.0\\
00031766056	& 06/13		& 4.4		& $1.9 \pm 0.2$			& 61			& 1.5\\
00031766057	& 06/19		& 5.0		& $1.3 \pm 0.2$			& 47			& 2.2\\
00031766058	& 06/25		& 4.2		& $2.1 \pm 0.3$			& 67			& 1.7\\
00031766059	& 07/01		& 4.4		& $2.5 \pm 0.3$			& 83			& 1.5\\
00031766060	& 07/09		& 5.0		& $1.6 \pm 0.2$			& 60			& 2.3\\
00031766061	& 07/13		& 4.2		& $2.2 \pm 0.3$			& 71			& 1.2\\
00031766062	& 07/19		& 4.6		& $8.4 \pm 0.5$			& 325			& 1.1\\
00031766063	& 07/25		& 4.3		& $13.1 \pm 0.7$		& 500			& 1.1\\
00031766064	& 08/06		& 5.0		& $2.1 \pm 0.2$			& 79			& 2.3\\
00031766065	& 08/12		& 5.0		& $2.0 \pm 0.2$			& 76			& 2.0\\
00031766066	& 08/18		& 5.4		& $2.5 \pm 0.3$			& 107			& 2.7\\
00031766067	& 08/24		& 4.9		& $2.2 \pm 0.2$			& 82			& 1.5\\
00031766068	& 08/30		& 4.7		& $1.7 \pm 0.2$			& 64			& 2.9\\
00031766069	& 09/05		& 4.2		& $2.2 \pm 0.3$			& 67			& 2.8\\
00031766070	& 09/11		& 4.2		& $2.3 \pm 0.3$			& 74			& 1.9\\
00031766071	& 09/17		& 5.1		& $2.3 \pm 0.3$			& 89			& 2.2\\
00031766072	& 09/23		& 1.8		& $2.0 \pm 0.4$			& 27			& 1.5\\
00031766073	& 09/29		& 4.6		& $1.8 \pm 0.2$			& 61			& 2.6\\
00031766074	& 10/05		& 4.5		& $1.4 \pm 0.2$			& 49			& 2.2\\
00031766075	& 10/11		& 3.3		& $1.2 \pm 0.2$			& 32			& 2.7\\
00031766076	& 10/16		& 4.4		& $1.2 \pm 0.2$			& 40			& 2.8\\
00031766077	& 10/17		& 4.9		& $1.3 \pm 0.2$			& 49			& 2.7\\
00031766078	& 10/23		& 4.7		& $1.2 \pm 0.2$			& 44			& 5.0\\
00031766079	& 10/29		& 4.8		& $1.4 \pm 0.2$			& 52			& 2.5\\
00031766080	& 11/04		& 4.7		& $1.2 \pm 0.2$			& 43			& 2.8\\
00031766081	& 11/09		& 4.5		& $1.1 \pm 0.2$			& 40			& 2.9\\
\hline
\end{tabular}
\end{center}
\label{tab:Photon Counting}
\end{table*}

X--ray data in the 0.3--10 keV band were collected by the XRT in the photon counting (PC) 
mode (2.5-s time resolution). We reduced and analysed the data with the HEASOFT software
package (version 6.13), as discussed in the {\it Swift}/XRT data reduction 
guide\footnote{\url{http://www.swift.ac.uk/analysis/xrt/index.php}}. 
We processed the data with the default parameter settings and created exposure maps for 
each event file with the {\tt xrtpipeline} task (v. 0.12.6). We determined the count rates 
through the ``detect'' command of {\tt ximage} (v. 4.5.1). All count rates values are well 
below $0.5$ c s$^{-1}$, providing that X--ray data are unaffected by pile-up ($0.5$ c s$^{-1}$ 
is actually the limiting value in the PC mode at which this effect should be checked).
We extracted all the source spectra with {\tt xselect} (v. 2.4b), adopting a circle centered 
on the source and of radius 10 pixels (15--20 pixels when the flux is higher) as extraction 
region. We then extracted the corresponding background spectra through annuli centered on the 
source with inner and outer radius of 40 and 80 pixels, respectively. We generated the ancillary
response files (ARFs) for each extracted spectrum using the {\tt xrtmkarf} task (v. 0.6.0), in 
order to correct the net count rate of each observation for losses due to the finite size of the
extraction region compared to the point spread function, vignetting and hot columns. We then 
assigned the latest version of the redistribution matrix files found in the CALDB database and 
we grouped all the spectra in the 0.3--10 keV energy range to a minimum of 20 counts per bin, in
order to ensure the applicability of the ${\chi}^{2}$ test. Single consecutive observations 
characterised by a low number of counts were grouped together to improve the fit statistics, for 
a total of 17 sets of XRT observations. ARFs were summed with the {\tt addarf} task after been 
weighted for the fractional number of counts of the corresponding observation with respect to the
total number of counts of the grouped observations.

Aql X-1 is not detected in none of the {\it Swift}/UVOT available data. A detailed study of the
UV/optical counterpart of this source in quiescence was not carried out in light of the presence 
of the nearby interloper, which would make practically impossible any significant quantitative analysis.

\begin{figure}
\begin{center}
\includegraphics[width=5.5cm,angle=-90]{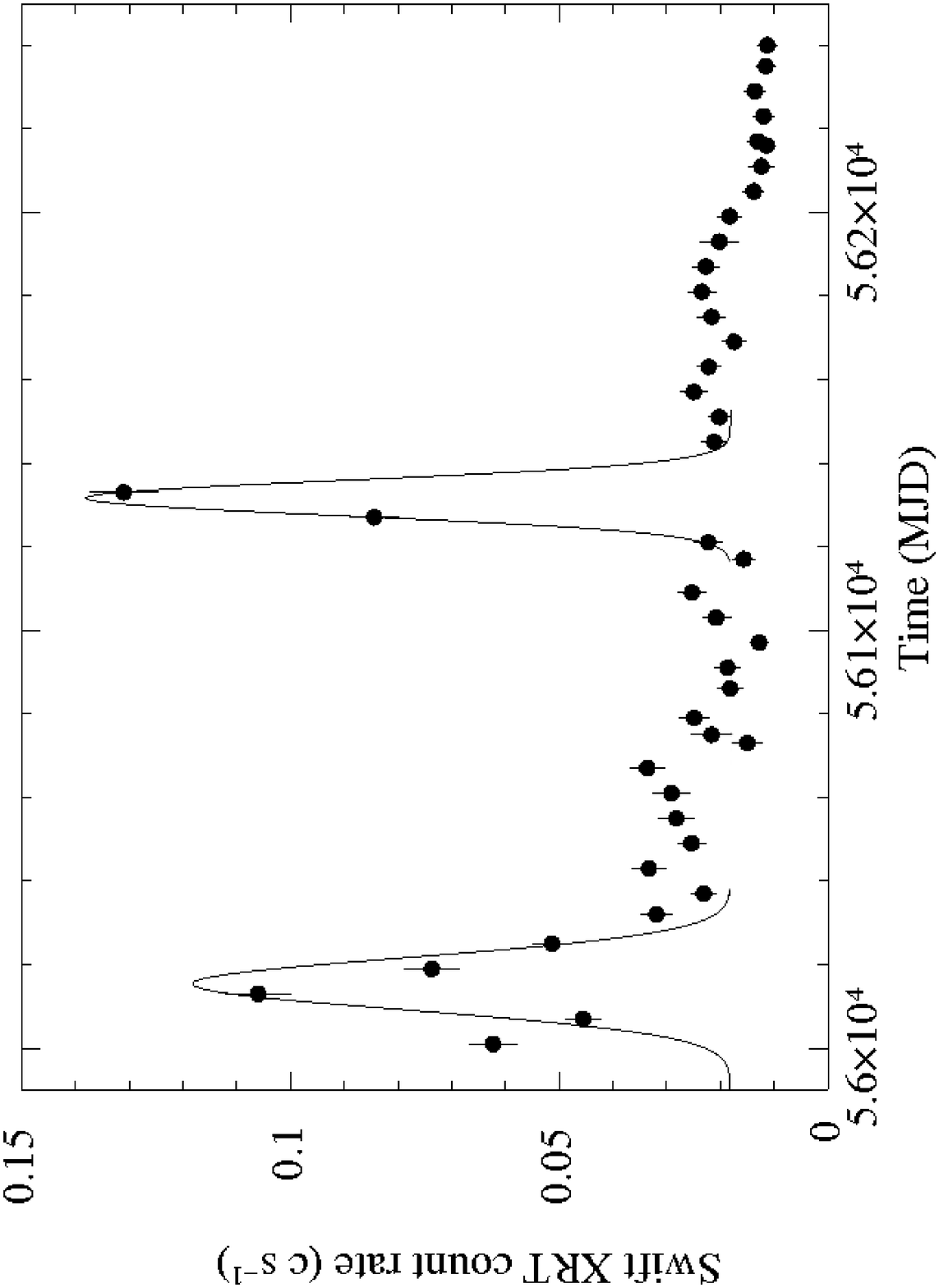}
\end{center}
\caption{0.3--10 keV light curve of Aql X-1 during the quiescent phase observed by {\it Swift}/XRT (15 March--9 November 
2012). Bin time equals one single observation.}
\label{fig:2012quiesc}
\vskip -0.1truecm
\end{figure}

\begin{figure}
\begin{center}
\includegraphics[width=5.5cm,angle=-90]{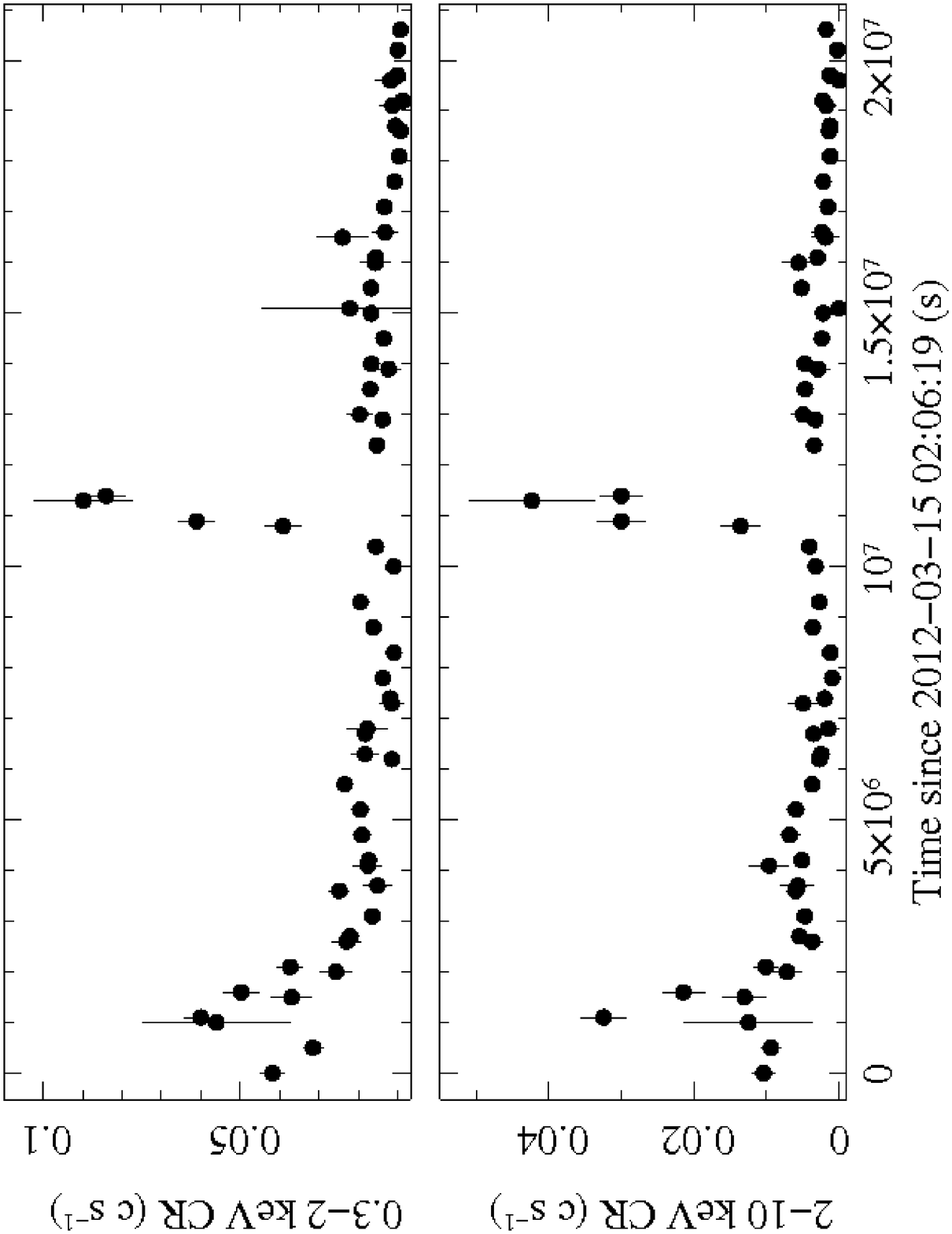}
\end{center}
\caption{Light curve of Aql X-1 in quiescence in the 0.3--2 keV ({\it top}) and 2--10 keV ({\it bottom}) energy bands. 
All the observations are grouped with a bin time of 10$^5$ s.}
\label{fig:cratesintervals}
\vskip -0.1truecm
\end{figure}

\begin{figure}
\begin{center}
\includegraphics[width=5.5cm,angle=-90]{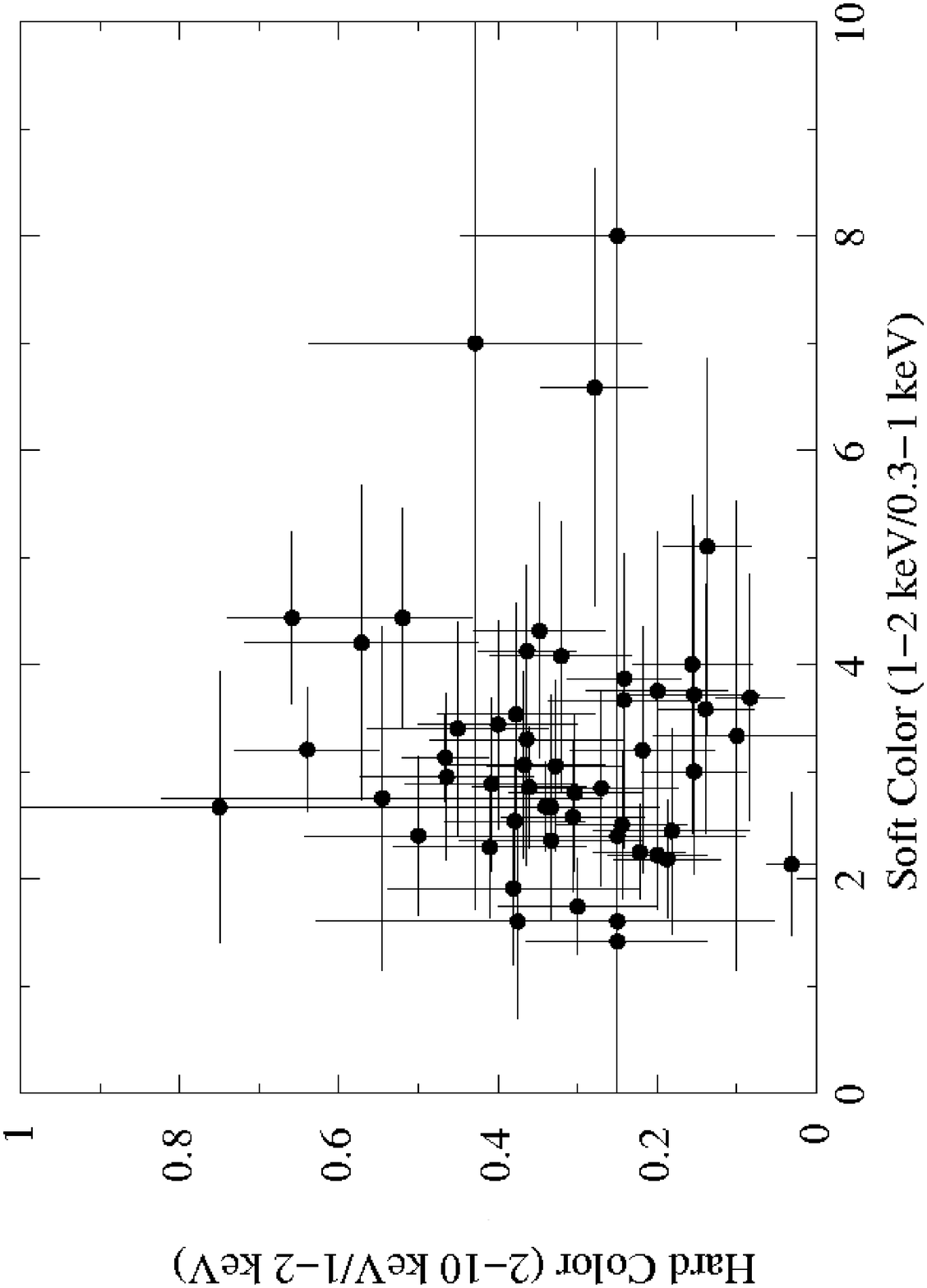}
\end{center}
\caption{Color-color diagram of Aql X-1 in quiescence. All the observations are grouped with a bin time of 10$^5$ s.}
\label{fig:colorcolor}
\vskip -0.1truecm
\end{figure}

\begin{figure}
\begin{center}
\includegraphics[width=5.5cm,angle=-90]{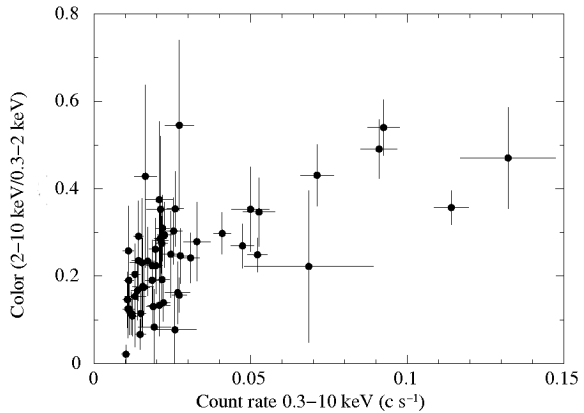}
\end{center}
\caption{Color-intensity diagram of Aql X-1 in quiescence. All the observations are grouped with a bin time of 10$^5$ s.}
\label{fig:colorintensity}
\vskip -0.1truecm
\end{figure}

\section{Data analysis and results}
Aql X-1 is highly variable during the 8 months of quiescence observed by 
{\it Swift}/XRT in 2012 (see Fig. 1 and Fig. 2 for the X--ray light curves).
The average count rate is $[2.92\pm0.07]\times10^{-2}$ c s$^{-1}$ and
the RMS fractional variation is found to be $0.84\pm0.08$ ({\tt lcstats} v. 1.0).
The source is more intense in the first 6 observations (Obs. 00031766040-45, 
March 15--April 15) than in the following 16 observations (Obs. 00031766046-61; 
April 20--July 13), when the detected count rate decreases by a factor of $\sim$1.7 
on average. This feature is probably due to the occurrence of a flare like event. 
In the second half of July, $\sim$3.5 months after the beginning of the monitoring 
program, the source showed another flare, with a peak count rate of 0.13 c s$^{-1}$ 
(Obs. 00031766063; July 25). During the remaining observations (Obs. 00031766064-81, 
August 6--November 9), the count rate is $\leq2.5\times10^{-2}$ c s$^{-1}$ and the 
X--ray light curve is almost flat (the RMS fractional variation is $< 0.11$ at a 3$\sigma$
confidence level). We characterised the flares in terms of energetic, 
peak luminosity and duration by fitting the light curve data with gaussian models, 
as shown in Fig. 1. We determined the number of counts by numerically integrating 
the model and the peak count rate and the duration by estrapolation of the
model. We considered the width of the gaussian as a rough estimate for the duration. 
The first flare releases an energy of $\sim6\times10^{40}$ erg, reaches a peak 
luminosity (for a source distance of 5 kpc) of $\sim4\times10^{34}\ergs$ in the 
0.3--10 keV energy band and lasts $\sim$1 month, while the second one releases an 
energy of $\sim4\times10^{40}$ erg, reaches a peak luminosity of $\sim4\times10^{34}\ergs$ 
in the same band and lasts $\sim$2 weeks. 

To study the quiescent spectral changes of Aql X-1 independently of the spectral models, 
we constructed the color-color and the color-intensity diagrams with {\tt lcurve} 
(bin time of 10$^5$ s). We defined the color as the ratio between the number of counts 
in the 2--10 keV and in the 0.3--2 keV ranges, and the hard and soft colors as the ratio
of counts  (2--10 keV/1--2 keV) and (1--2 keV/0.3--1 keV), respectively. The intensity 
refers to the count rate in the 0.3--10 keV band. The color-color diagram (see Fig. 3) 
is suggestive of quiescent variability and the color-intensity diagram (see Fig. 4) reveals 
that the spectrum is softer at lower intensity (count rate $\lsim0.08$ c s$^{-1}$) and becomes
harder as the intensity increases (count rate $\gsim0.08$ c s$^{-1}$; i.e. during the flare).

\subsection{X--ray spectra}
We performed a detailed analysis of the 17 X--ray spectra of Aql X-1 with the {\tt Xspec} 
(v. 12.8.0) spectral fitting package (Arnaud 1996). We included the effects of interstellar 
absorption along the line of sight through the {\tt TBABS} model with the {\tt vern} cross 
sections (Verner et al. 1996) and {\tt wilm} abundances (Wilms, Allen \& McCray 2000). 
We adopted the {\tt NSATMOS} model (Heinke et al. 2006) for the NS atmosphere. This model 
assumes a negligible ($<10^9$ G) magnetic field and a static atmosphere of pure, ionised hydrogen.
It also incorporates thermal electron conduction and self-irradiation by photons from the 
NS. Its parameters are the effective temperature in the NS frame (expressed as 
Log$[T_{\rm eff}$/\deg K]), the NS mass ($M_{\rm NS}$) and radius ($R_{\rm NS}$), the
source distance ($D$) and a normalisation factor, which parametrizes the fraction of the
NS surface that is emitting. In all the fits we fixed $M_{\rm NS}$ and $R_{\rm NS}$ to 
the canonical NS values ($1.4\msole$ and 10 km, respectively), the source distance to 5
kpc and the normalisation to 1 (i.e. we assumed that the entire NS surface is radiating). 
Therefore, the effective temperature is the only parameter allowed to vary. Initially, we 
fit all the 17 spectra individually in the 0.3--10 keV range with three different models: 
{\tt NSATMOS}, power law and {\tt NSATMOS} plus power law, to account for both the soft 
and hard components. Despite the fact that the power law model is statistically the best
in some cases ($\chi^2_\nu\leq1.22$ for 5 out of 17 fits), the majority of the spectra are 
well fit by the two-component model ($\chi^2_\nu\leq1.14$ for 12 out of 17 fits; 
$1.32\leq\chi^2_\nu\leq1.71$ for the remaining 5 fits). In most of those fits however 
the values ​​of the parameters are poorly constrained. For this reason we further grouped 
the XRT observations, separating the flares (Obs. 00031766040-45 and Obs. 00031766062-63) 
from the pure quiescent phases (Obs. 00031766046-61 and Obs. 00031766064-81), for a total 
of four data sets. We fit the corresponding spectra with the {\tt NSATMOS} plus power law 
model (see Fig. 5 and Fig. 6 for examples). Best fitting parameters are given 
in Table 2. All uncertainties were determined with the {\tt steppar} command in {\tt Xspec}
and are quoted at the 90\% confidence level for a single interesting parameter ($\Delta\chi^2=2.706$). 
The fit values for the column density are consistent with being constant within the errors. 
The NS effective temperature (as measured by an observer at infinity) during the first flare 
($kT_{\rm eff}^\infty=123.6^{+8.6}_{-17.1}$ eV) is only slightly larger than the one 
relative to the following phase ($kT_{\rm eff}^\infty=120.7^{+1.9}_{-2.8}$ eV). Due to 
the relatively low number of counts, only an upper limit of $kT_{\rm eff}^\infty\lsim148.5$
eV is inferred during the second flare and after that the temperature decreases to
$kT_{\rm eff}^\infty=107.6^{+2.5}_{-7.4}$ eV. The power law photon index ($\Gamma$) is more 
or less similar during the two flares ($2.0\pm0.5$ and $2.5^{+0.5}_{-0.9}$, respectively) 
and only an upper limit can be derived during the quiescent phases ($\lsim1.4$ and $\lsim2.5$,
respectively). The fractional contribution of the power law component to the total unabsorbed 
flux in the 0.3--10 keV energy range is larger during flares ($\sim53$\% and $\sim56$\%, 
respectively) than during pure quiescence ($\sim34$\% and $\sim22$\%, respectively).

\begin{figure}
\begin{center}
\includegraphics[width=5.5cm,angle=-90]{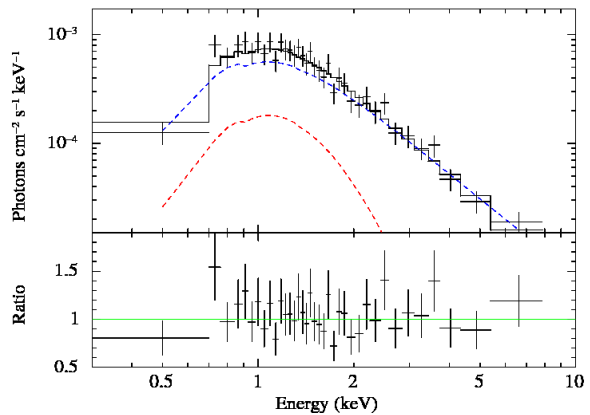}
\end{center}
\caption{X--ray spectrum of Aql X-1 relative to the second flare (Obs. 00031766062-63). 
	 Dashed lines refer to the {\tt NSATMOS} (red) and power law (blue) unfolded spectra.}
\label{fig:flarespectrum}
\vskip -0.1truecm
\end{figure}

\begin{figure}
\begin{center}
\includegraphics[width=5.5cm,angle=-90]{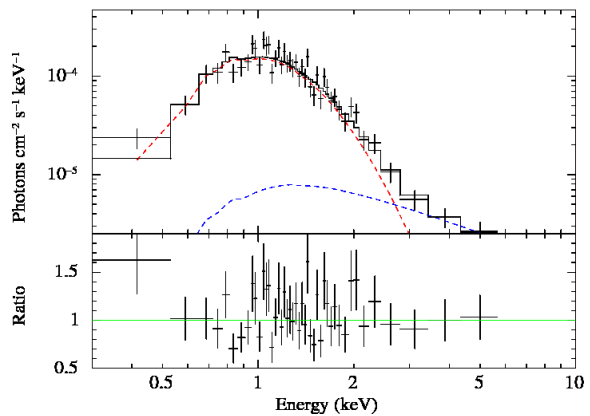}
\end{center}
\caption{X--ray spectrum of Aql X-1 relative to the second purely quiescent phase (Obs. 00031766064-81). 
	 Dashed lines refer to the {\tt NSATMOS} (red) and power law (blue) unfolded spectra.}
\label{fig:quiescentspectrum}
\vskip -0.1truecm
\end{figure}

\begin{table*}
\footnotesize
\caption{Spectral fitting parameters. The spectra were fit with the {\tt NSATMOS} plus power law model. We fixed the NS 
mass and radius to the canonical values ($1.4 \msole$ and 10 km, respectively) and the {\tt NSATMOS} normalisation to 1.}
\begin{threeparttable}
\begin{tabular}{ccccccccc}
\hline
Obs. ID 	     & \nh	 		& $kT_{\rm eff}^\infty$\tnote{a}& $\Gamma$		& PL norm.\tnote{b}		& $L$\tnote{c}			& PL fraction\tnote{d} 	& $\chi^2_\nu$ (d.o.f.)	& NHP	\\ 
		     & (10$^{21}$ cm$^{-2}$)  	& (eV)		   	        & 		  	& (10$^{-4}$)			& (10$^{33} \ergs$)		& 			&			& (\%)	\\ 
\hline
\vspace{0.08cm}  
00031766040-45	     & $5.6_{-0.6}^{+0.9}$	& $123.6_{-17.1}^{+8.6}$     	& $2.0 \pm 0.5$ 	& $3.5_{-2.0}^{+4.0}$		& $14.6_{-6.4}^{+12.5}$		& 0.53 			& 1.20 (53)		& 15.4	\\ \vspace{0.08cm} 
00031766046-61	     & $6.5 \pm 0.6$ 		& $120.7_{-2.8}^{+1.9}$		& $\lsim 1.4$		& $0.06_{-0.01}^{+0.3}$		& $5.1_{-0.8}^{+1.9}$		& 0.34			& 0.85 (45) 		& 75.8	\\ \vspace{0.08cm}
00031766062-63	     & $6.3_{-1.7}^{+1.6}$	& $\lsim 148.5$     		& $2.5_{-0.9}^{+0.5}$	& $16.6_{-12.7}^{+12.1}$	& $35.9_{-4.1}^{+8.3}$		& 0.56			& 0.70 (33)		& 90.0	\\ \vspace{0.08cm}
00031766064-81	     & $5.2_{-0.5}^{+0.6}$	& $107.6_{-7.4}^{+2.5}$		& $\lsim 2.5$		& $0.2_{-0.1}^{+0.9}$		& $2.9_{-0.6}^{+0.7}$		& 0.22			& 1.11 (41)		& 28.7	\\ 
\hline
\end{tabular}
\label{tab:spectralfits}
\begin{tablenotes}
\item[a] The effective temperature as measured by an observer at infinity is $kT_{\rm{eff}}^{\infty}= kT_{\rm{eff}}/(1+z)$, 
where $1+z = (1-R_{\rm{s}}/R_{\rm{NS}})^{-1/2}$ is the gravitational redshift factor ($R_{\rm{s}}=2GM_{\rm{NS}}/c^2$ is the 
Schwarzschild radius, $G$ the gravitational constant and $c$ the speed of light).
\item[b] The power law normalisation is in units of photons keV$^{-1}$ cm$^{-2}$ s$^{-1}$ and is defined at 1 keV.
\item[c] The luminosity is evaluated over the 0.3--10 keV range and for an assumed distance of $D=5$ kpc.
\item[d] Fractional contribution of the power law component to the total unabsorbed flux in the 0.3--10 keV energy range.
\end{tablenotes}
\end{threeparttable}
\end{table*}

\subsection{Simultaneous spectral fitting}
In order to physically constrain the powering mechanism(s) at work in quiescence, we fit 
all 17 spectra simultaneously with the {\tt NSATMOS} plus power law model and with different 
combinations of parameters tied and allowed to vary between the observations (10 simultaneous
fits in total). We summarise the results in Table 3. F-test probabilities, which are 
determined through a comparison with the case where only the power law normalisation is allowed
to vary, are also reported. Differences between spectra cannot be attributed to a variable column
density alone ($\chi^2_\nu=5.56$), a variable thermal component alone ($\chi^2_\nu=1.80$),
or to changes in both these quantities ($\chi^2_\nu=1.82$). Six different models provide a 
statistically good fit and the F-test shows that no model with two or more variable parameters is 
statistically better than the one where only the power law normalisation is allowed to vary 
($\chi^2_\nu = 0.97$, 175 degrees of freedom, null hypothesis probability $\sim59\%$). 
Spectral fitting parameters obtained with this model are given in Table 4. We found a hydrogen 
column density $\nh=5.9_{-0.4}^{+0.3}\times10^{21}$ cm$^{-2}$, an effective temperature 
$kT_{\rm eff}^\infty = 105.2_{-2.2}^{+1.7}$ eV and a power law photon index $\Gamma=2.4\pm0.1$. 
The power law normalisation varies by a factor $\sim80$ between the pure quiescent observations 
($\lsim0.3\times10^{-4}$ keV$^{-1}$ cm$^{-2}$ s$^{-1}$; Obs. 00031766075-78 and 00031766079-81) 
and the one relative to the peak of the flare ($20.6_{-2.7}^{+3.0}\times10^{-4}$ keV$^{-1}$ 
cm$^{-2}$ s$^{-1}$; Obs. 00031766063). Both the thermal and non-thermal fluxes increase 
during the flares and the corresponding spectra show an enhanced contribution of the power 
law tail to the total unabsorbed flux in the 0.3--10 keV range. We show the temporal evolution of
the power law normalisation and fraction in Fig. 7.

\begin{table*}
\scriptsize
\caption{Results of the simultaneous spectral fits with the {\tt NSATMOS} plus power law model. We adopted the same 
prescriptions as above for the {\tt NSATMOS} model. F-test probabilities are determined through a comparison with the
case where only the power law normalisation is allowed to vary.}
\begin{threeparttable}
\begin{tabular}{llcc}
\hline
\vspace{0.08cm}
 Tied parameters 				& Variable parameters  	 			& $\chi^2_\nu$ (d.o.f.) & F-test probability (\%)\\ 
 \hline 
 \vspace{0.08cm}
 \nh, $kT_{\rm eff}^\infty$, $\Gamma$, PL norm	& -	      		 			& 7.05 (191)		& -			\\ \vspace{0.08cm} 
 $kT_{\rm eff}^\infty$, $\Gamma$, PL norm	& \nh 	      	 				& 5.56 (175)		& -			\\ \vspace{0.08cm} 
 \nh, $\Gamma$, PL norm   			& $kT_{\rm eff}^\infty$	      	 		& 1.80 (175)		& -			\\ \vspace{0.08cm}
 \nh, $kT_{\rm eff}^\infty$, $\Gamma$		& PL norm  	      	 			& 0.97 (175)		& -			\\ \vspace{0.08cm}
 $\Gamma$, PL norm 				& \nh, $kT_{\rm eff}^\infty$ 	 		& 1.82 (159)		& -			\\ \vspace{0.08cm}
 \nh, $\Gamma$  				& $kT_{\rm eff}^\infty$, PL norm 	 	& 0.90 (159)		& 2.89			\\ \vspace{0.08cm}
 \nh, $kT_{\rm eff}^\infty$  			& $\Gamma$, PL norm	 			& 0.90 (159)		& 2.89			\\ \vspace{0.08cm}
 $\Gamma$	  				& \nh, $kT_{\rm eff}^\infty$, PL norm 		& 0.87 (143)		& 2.84			\\ \vspace{0.08cm}
 $kT_{\rm eff}^\infty$	  			& \nh, $\Gamma$, PL norm 			& 0.90 (143)		& 8.34			\\ \vspace{0.08cm}
 \nh		  				& $kT_{\rm eff}^\infty$, $\Gamma$, PL norm 	& 0.89 (143)		& 5.95			\\ 
\hline
\end{tabular}
\label{tab:simultaneousspectralfits}
\end{threeparttable}
\end{table*}

\begin{table*}
\footnotesize
\caption{Spectral fitting parameters. The spectra were fit simultaneously with the {\tt NSATMOS} plus power law model
and with the column density, the NS effective temperature and the power law photon index tied between all the spectra.}
\begin{center}
\begin{tabular}{lclclcccc}
\hline
Obs. ID		& $N_{\rm H}$ 		& $kT_{\rm eff}^{\infty}$ 	& $\Gamma$ 	& PL norm. 		& $L$ 			   	& PL	 	& $\chi_\nu^2$ (d.o.f.)	& NHP  \\
		& ($10^{21}$ cm$^{-2}$) & (eV) 				& 		& ($10^{-4}$) 		& ($10^{33}$ erg s$^{-1}$) 	& fraction 	&			& (\%)\\ 
\hline \\ \vspace{0.06cm}
00031766040 	& $5.9^{+0.3}_{-0.4}$ 	& $105.2^{+1.7}_{-2.2}$ 	& $2.4\pm0.1$ 	& $7.5^{+1.5}_{-1.3}$ 	& 15.1 				& 0.41  	& 0.97 (175)		& 59 	\\ \vspace{0.06cm}
00031766041	& 			& 				& 		& $5.7^{+1.2}_{-1.1}$ 	& 12.1 				& 0.56  	& 			&	\\ \vspace{0.06cm}
00031766042  	& 			& 				& 		& $13.8^{+2.2}_{-1.9}$ 	& 23.5 				& 0.75 		& 			&	\\ \vspace{0.06cm}
00031766043  	&  			&				& 		& $9.5^{+1.7}_{-1.5}$ 	& 17.4 				& 0.67 		& 			&	\\ \vspace{0.06cm}
00031766044  	&  			& 				& 		& $5.4^{+1.2}_{-1.0}$ 	& 12.2 				& 0.38 		& 			&	\\ \vspace{0.06cm}
00031766045-46  &  			& 				& 		& $2.2\pm0.9$ 		& 6.7 				& 0.35 		&			&	\\ \vspace{0.06cm}
00031766047-48  &  			& 				& 		& $1.6\pm1.0$ 		& 7.1 				& 0.61 		& 			&	\\ \vspace{0.06cm}
00031766049-51  &  			& 				& 		& $2.9^{+0.7}_{-0.6}$ 	& 7.7 				& 0.35 		& 			&	\\ \vspace{0.06cm}
00031766052-56  &  			& 				& 		& $1.0\pm0.4$ 		& 4.9 				& 0.23 		& 			&	\\ \vspace{0.06cm}
00031766057-61  &  			& 				& 		& $1.2^{+0.4}_{-0.3}$ 	& 5.1 				& 0.24 		& 			&	\\ \vspace{0.06cm}
00031766062  	&  			& 				& 		& $13.0^{+2.0}_{-1.8}$ 	& 22.5 				& 0.70 		& 			&	\\ \vspace{0.06cm}
00031766063  	&  			& 				& 		& $20.6^{+3.0}_{-2.7}$ 	& 34.7 				& 0.55 		& 			&	\\ \vspace{0.06cm}
00031766064-66  &  			& 				& 		& $1.7^{+0.5}_{-0.4}$ 	& 5.8 				& 0.32 		& 			&	\\ \vspace{0.06cm}
00031766067-70  &  			& 				& 		& $1.5\pm0.5$	 	& 5.5	 			& 0.41		& 			&	\\ \vspace{0.06cm}
00031766071-74  &  			& 				& 		& $0.9^{+0.4}_{-0.3}$	& 4.7	 			& 0.23		& 			&	\\ \vspace{0.06cm}
00031766075-78  &  			& 				& 		& $\lsim 0.3$	 	& 3.7	 			& 0.19		& 			&	\\ \vspace{0.06cm}
00031766079-81  &  			& 				& 		& $\lsim 0.3$	 	& 4.1	 			& 0.23		& 			&	\\ 
\hline
\end{tabular}
\end{center}
\label{tab:specfit}
\end{table*}

\begin{figure}
\begin{center}
\includegraphics[width=6cm,angle=-90]{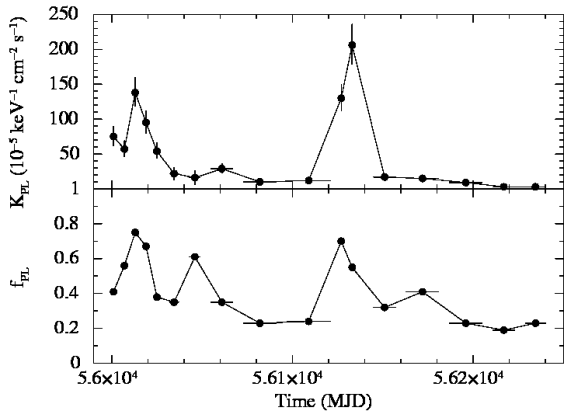}
\end{center}
\caption{Evolution of the power law normalisation ($K_{PL}$, {\it top}) and fraction ($f_{PL}$, 
{\it bottom}) as a function of time in the two-component spectral model where only the power law 
normalisation is allowed to vary between the observations.}
\label{fig:Kplflux}
\vskip -0.1truecm
\end{figure}

\section{Discussion and conclusions}
The most detailed observational campaign ever of the NS-LMXT Aql X-1 quiescent phase has been realised 
in the Mar-Nov 2012 interval with the {\it Swift} satellite. The source is found to be clearly variable 
on time-scale of weeks. Two flares were detected, with approximately comparable peak luminosities
($\sim4\times10^{34}\ergs$ in the 0.3--10 keV energy band), but different durations ($\sim$1 month and 
$\sim$2 weeks, respectively). 

Spectral data are best described by a model consisting of a soft thermal component (a neutron star 
atmosphere model) below $\sim$2 keV plus a hard non-thermal component (a power law tail) above $\sim$2 
keV. The non-thermal component increases faster than the thermal component during the flares, leading 
to a power law fraction that is higher during flares than during purely quiescent periods. A few flares
have been reported to occur in other NS-LMXTs in quiescence: regular observations of the Galactic centre
with {\it Swift}/XRT revealed a short flare from GRS J1741-2853, lasting $\sim1$ week and with a peak 
luminosity of $\sim9\times10^{34}$ erg s$^{-1}$ in the 2--10 keV range. The same observational campaign 
detected another flare from KS 1741-293, lasting $\lsim4$ days and with a peak luminosity of 
$\sim3\times10^{34}$ erg s$^{-1}$ in the 2--10 keV range (Degenaar \& Wijnands 2009, 2013). XTE J1701-462
showed several flares with peak luminosities in the $10^{34-35}$ erg s$^{-1}$ range (Fridriksson et al. 
2011) and more recently a flare from SAX J1750.8-2900 was detected, with a duration $\lsim16$ days and 
peak luminosity of $\sim3-4\times 10^{34}$ erg s$^{-1}$ in the 0.5--10 keV range (Wijnands \& Degenaar 2013).
All these episodes have been interpreted in terms of episodic low-level accretion activity during quiescence. 
This may be the case also for Aql X-1. A relatively steep power law is required when fitting the spectra 
and the temperature is well constrained only for the first flare. This is quite similar to what has been 
found for the flare of SAX J1750.8-2900 (Wijnands \& Degenaar 2013). Its spectrum is best described by an
absorbed power law model ($\chi_\nu^2=1.30$ for 8 d.o.f.) with a relatively high photon index despite a 
large error ($\Gamma=2.4\pm1.3$). When a thermal component is added to this model, a good fit is obtained
but the low statistical quality of the data do not allow to put meaningful constraints on this component.

Based on our data set, the two-component spectral model which best describes the overall quiescent 
X--ray variability of Aql X-1 is the one where the power law normalisation is the only parameter
allowed to vary between the observations ($\chi_\nu^2=0.97$ for 175 d.o.f.; the F-test shows that 
no model with two or more variable parameters is statistically better). This model suggests that 
the quiescent X--ray emission of Aql X-1 is set by a soft thermal component that is stable over 
time-scales of months and by a variable power law tail at higher energies. 
The cooling of the neutron star core might be the dominant emission source of the thermal emission 
and different mechanisms involving mass accretion onto the neutron star surface or magnetosphere 
might be responsible for the variability of the hard component. In fact, the pulsar scenario is 
disfavoured in justifying the overall variability, since it can hardly be reconciled with the large
and rapid variations of the power law normalisation observed during the flares, $\sim$2 orders of 
magnitudes in a few weeks (see Campana \& Stella 2003). We note however that we cannot rule out from
our analysis that the thermal component can be varying as well, since also this model gives a 
statistically acceptable result ($\chi_\nu^2=0.90$ for 159 d.o.f.).

Cackett et al. (2011) analysed 14 archival observations of Aql X-1 in quiescence.
These observations were performed with different satellites (11 with {\it Chandra}
and 3 with {\it XMM-Newton}) between November 2000 and October 2002, and Aql X-1 went
in outburst twice during those years (Campana et al. 2013).
The authors found variability between the observations and fit all the spectra 
simultaneously with the same two-component model in order to determine which spectral 
parameters varied. Allowing the power law normalisation alone to vary gives a good
fit ($\chi_\nu^2=1.10$ for 533 d.o.f.), as well as allowing both this parameter
and the neutron star effective temperature to vary gives acceptable results 
($\chi_\nu^2=0.88$ for 520 d.o.f.). Therefore, the authors were unable to conclusively 
determine whether the power law and/or the thermal component was driving the variability 
(see also Rutledge et al. 2002; Campana \& Stella 2003).

Quiescent variability of Cen X-4 has been revealed with {\it Swift} on time-scale of days
between June and August 2012 (Bernardini et al. 2013). A total of 60 spectra were grouped
according to their count rate in three ranges and then summed. The same two-component model
was used to fit the three spectra. Spectral changes cannot be accounted for neither by a 
change in the column density alone ($\chi_\nu^2=2.10$ for 193 d.o.f.), nor by changes 
in the column density plus the power law component ($\chi_\nu^2=1.25$ for 191 d.o.f.)
or the thermal component ($\chi_\nu^2=1.52$ for 193 d.o.f.). The best model is instead
the one where both the power law and the neutron star effective temperature are free to 
vary ($\chi_\nu^2=1.00$ for 189 d.o.f.). Moreover, the X--ray spectral shape remained
approximately constant independently of flux variations, implying that both the soft and
hard components varied in tandem with the X--ray flux. To explain the close link between
the two spectral components, it was suggested that the quiescent X--ray emission of 
Cen X-4 is produced by accretion onto the NS surface. Interestingly, this link has 
been observed also on time-scales of years (Cackett et al. 2010).

Unlike Cen X-4, a tight connection between the two spectral components is not observed in
Aql X-1, neither on time-scales of weeks (the variability can be accounted for by changes of 
the power law alone), nor of years (Cackett et al. 2011). This possibly different behaviour
may lie in the intrinsically different nature of these two sources. In fact, only two outbursts 
have been detected from Cen X-4 since its discovery in 1969, with the last one occurring $\sim34$ 
yr ago (Kaluzienski, Holt \& Swank 1980). On the other hand, Aql X-1 is the most prolific NS-LMXT, 
showing outbursts with a recurrence time of $280\pm103$ d at $1\sigma$ level in the last 
$\sim20$ yr (Campana et al. 2013). Moreover, no exact determinations of the spin period 
of Cen X-4 exist and thus, assuming that the magnetic field corotates with the neutron star, 
no accurate predictions about the effects of the magnetospheric centrifugal drag on the 
accreting matter can be formulated for this source. One can speculate that different spin 
periods may imply a different dynamics of mass accretion.

More frequent observations of the quiescence of Aql X-1 with active X-ray observatories 
({\it Swift}, {\it XMM-Newton} and {\it Chandra}) may help to cast new light on its strong X--ray
variability on shorter time-scales. 

\section*{Acknowledgments}
We thank the referee and Federico Bernardini for useful comments and suggestions.

\end{document}